%% file: EquilibriumMeasures_revision.tex
\pdfoutput=1
\documentclass{article}
\usepackage{epsfig}
\usepackage{amssymb}
\usepackage{amsbsy}
\usepackage{wrapfig}
\usepackage{color}
\usepackage{graphicx}
\def \LN{\ln \frac 1{|z-t|}}

\def\supp{\mathrm{supp}}

\usepackage{verbatim}
\def\wt{\widetilde}

\def\wh{\widehat}

\def\res{\mathop{\mathrm{res}}\limits_}

\makeatletter
\@addtoreset{equation}{section}
\makeatother

\def\le{\left}
\def\ri{\right}

\def\bc{\begin{corollary}}
\def\ec{\end{corollary}}
\def\&{&{\hskip -20pt}}

\def\br{\begin{remark}\rm\small}
\def\1{{\bf 1}}
\def\er{\end{remark}}
\def\bt{\begin{theorem}}
\def\et{\end{theorem}}

\newcommand{\M}{\ensuremath{\mathcal M}}
\newcommand{\F}{\ensuremath{\mathcal F}}
\newcommand{\V}{\ensuremath{\mathcal V}}
\newtheorem{corollary}{Corollary}[section]

\def\bx{\begin{example}}
\def\ex{\end{example}}
\def\bd{\begin{definition}}
\def\ed{\end{definition}}
\def\bp{\begin{proposition}\rm}
\def\bl{\begin{lemma}\em}
\def\el{\end{lemma}}
\def\ep{\end{proposition}}
\def\be{\begin{equation}}
\def\ee{\end{equation}}
\def\bea{\begin{eqnarray}}
\def\eea{\end{eqnarray}}

\def\C{{\mathbb C}}
\def\R{{\mathbb R}}

\newtheorem{theorem}{Theorem}[section]

\newtheorem{coroll}{Corollary}[section]
\newtheorem{examps}{Examples}[section]

\newtheorem{lemma}{Lemma}[section]
\newtheorem{remark}{Remark}[section]
\newtheorem{remarks}[remark]{Remarks}
\newtheorem{proposition}{Proposition}[section] 
\newtheorem{definition}{Definition}[section]
\def\br{\begin{remark}}
\def\er{\end{remark}}
\def\bt{\begin{theorem}}
\def\et{\end{theorem}}
\def\bc{\begin{coroll}}
\def\ec{\end{coroll}}
\def\brs{\begin{remarks} \rm\
\begin{enumerate}}
\def\ers{\end{enumerate}\end{remarks}}
\def\bl{\begin{lemma}}
\def\el{\end{lemma}}
\def\bxs{\begin{examps}. \rm\begin{enumerate}}
\def\exs{\end{enumerate}\end{examps}}
\def\bd{\begin{definition}}
\def\ed{\end{definition}}
\def\bp{\begin{proposition}}
\def\ep{\end{proposition}}
\def\be{\begin{equation}}
\def\ee{\end{equation}}

\def\d{{\rm d}}
\def\bea{\begin{eqnarray}}
\def\eea{\end{eqnarray}}
\def\beas{\begin{eqnarray*}}
\def\eeas{\end{eqnarray*}}

\def\iint{\int\!\!\!\!\int}
\def\C{{\mathbb C}}

\def\R{{\mathbb R}}

\date{}
\begin{document}
\fontfamily{cmss}
\fontsize{11pt}{15pt}
\selectfont

\baselineskip 16pt plus 1pt minus 1pt
%\begin{titlepage}
\vspace{0.2cm}
\begin{center}
\begin{Large}\fontfamily{cmss}
\fontsize{17pt}{27pt}
\selectfont
\textbf{Regularity of a vector potential problem and its spectral curve}
\end{Large}\\
\bigskip
\begin{large}
{F. Balogh}$^{\dagger\ddagger}$\footnote{balogh@crm.umontreal.ca},
{M. Bertola}$^{\dagger\ddagger}$\footnote{Work supported in part by the Natural
    Sciences and Engineering Research Council of Canada (NSERC).}\footnote{bertola@crm.umontreal.ca}
\end{large}

\bigskip
\begin{small}
$^{\dagger}$ {\em Centre de recherches math\'ematiques,
Universit\'e de Montr\'eal\\ C.~P.~6128, succ. centre ville, Montr\'eal,
Qu\'ebec, Canada H3C 3J7} \\
\smallskip
$^{\ddagger}$ {\em Department of Mathematics and
Statistics, Concordia University\\ 1455 de Maisonneuve Blvd. West, Montr\'eal, Qu\'ebec,
Canada H3G 1M8} \\
\end{small}
\end{center}
\bigskip
%%%%%%%%%%%%%%%%%%%%%%%%%%%%%%%%%%%%%%%%%%%%%%%

%% Abstract

%%%%%%%%%%%%%%%%%%%%%%%%%%%%%%%%%%%%%%%%%%%%%%%
\begin{abstract}
In this note we study a minimization problem for a vector of measures subject to a prescribed interaction matrix in the presence of external potentials. The conductors are allowed to have zero distance from each other but the external potentials satisfy a growth condition near the common points.

We then specialize the setting to a specific problem on the real line which arises in the study of certain biorthogonal polynomials (studied elsewhere) and we prove that the equilibrium measures solve a pseudo--algebraic curve under the assumption that the potentials are real analytic. In particular the supports of the equilibrium measures are shown to consist of a finite union of compact intervals.
\end{abstract}

\medskip
%\begin{small}
%\end{small}
%\end{titlepage}
%\tableofcontents
%\newpage
%

%%%%%%%%%%%%%%%%%%%%%%%%%%%%%%%%%%%%%%%%%%%%%%%

%% Intro

%%%%%%%%%%%%%%%%%%%%%%%%%%%%%%%%%%%%%%%%%%%%%%%

\section{Introduction}
In this short paper we consider a vector potential problem of relevance in the study of the asymptotic behavior of multiple orthogonal polynomials for the so-called Nikishin systems \cite{Nikishin}. 
%\red
{The original problem was introduced in \cite{GoncharRakhmanovPade} (without external fields) and further questions has been addressed in \cite{GoncharRakhmanov,ns, GoncharRakhmanovSorokin,Aptekarev, Lapik}.} The main motivation of our interest for this problem arises in a recently introduced set of biorthogonal polynomials \cite{BertoGekhtSzmig1}. These polynomials are related on one side to the spectral theory of the ``cubic string'' and the DeGasperis--Procesi peakon solutions of the homonymous nonlinear differential equation \cite{dp}; on the other end they are related to a two--matrix model \cite{BertoGekhtSzmig2} with a measure of the form 
\be
\d\mu(M_1,M_2)  = \frac 1{\mathcal Z_N} \d M_1 \d M_2 \, \frac {\alpha(M_1)\beta(M_2)}{\det(M_1+M_2)^N}\label{12}
\ee
where the $M_j$'s are positive definite Hermitian matrices of size $N\times N$, $\alpha ,\beta$ are some positive densities on $\R_+$ and  the  expressions $\alpha(M_1), \beta(M_2)$ stand for the product of those densities on the spectra of $M_j$. 

The relation between the relevant biorthogonal polynomials and the above--mentioned matrix model is on the identical logical footing as the relation between ordinary orthogonal polynomials and the Hermitian random matrix model \cite{MehtaBook}.

In \cite{BertoGekhtSzmig1} a Riemann--Hilbert formulation (similar to the formulation of multiple--orthogonal polynomials as explained in \cite{VanAssche} but adapted to the peculiarities of the model)  was derived and in \cite{BertoGekhtSzmig2} the correlation functions of the spectra of the two matrices were completely characterized in terms of the matrix--solution of that Riemann--Hilbert problem.

In \cite{BertoGekhtSzmig3} the analysis of the  strong asymptotics with respect to varying weight (following \cite{DKMVZ}) will be carried out. A pre-requisite of that analysis is the existence and regularity of the solution of a suitable potential problem, namely the one which we explain in the second part of the paper.

In fact, the present paper is addressing a wider class of potential problems that will be necessary for the study of the spectral statistics in the limit of large sizes of the {\em multi--matrix model}
\be
\d\mu(M_1,\dots,M_R)  = \frac 1{\mathcal Z_N} \frac {\prod_{j=1}^{R} \alpha_j(M_j)\d M_j}{\prod_{j=1}^{R-1} \det(M_j + M_{j+1})^N}
\label{chain}
\ee 
corresponding to a {\em chain} of positive--definite Hermitian matrices $M_j$ with densities $\alpha_j$ as above.

In Section \ref{sect2} we consider the problem as a vector potential problem in the complex plane with a prescribed interaction matrix. Under a suitable growth condition for the external potentials $V_j(z)$ near the overlap region of the conductors (in particular the common points on the boundaries) it is shown that the minimizing vector of equilibrium measures has supports for the components separated by positive distances.

In Section \ref{sect3} we specialize the setting to  the situation in which the conductors $\Sigma_j = (-1)^{j-1} [0,\infty)$ (so that they have the origin in common), with an interaction matrix of Nikishin type as in \cite{VanAssche}. We prove that the minimizing measure is regular and supported in the interior of the condensers (under our assumption of growth of the potentials).

This result allows to proceed in Section \ref{sect4} with a manipulation of algebraic nature involving the Euler--Lagrange equations for the {\em resolvents} (Cauchy transforms) $W_j(x)$ of the  equilibrium measures. It is shown that certain auxiliary quantities $Z_j$  that depend linearly on the resolvents and the potentials (see (\ref{shiftresolvents}) for the precise formula) enter a pseudo--algebraic equation of the form 
\be
z^R + C_{2}(x) z^{R-1} + \dots + C_{R+1}(x)=0\label{introcurve}
\ee  
where the functions $C_j(x)$ are analytic functions with the same singularities as the derivative of the potentials $V'_k(x)$ in the common neighborhood of the real axis where all the potentials are real analytic. In particular the coefficients $C_j(x)$ do not have jumps on the real axis and the various branches of eq. \ref{introcurve} are precisely the $Z_j(x)$ defined above. For example, if the derivative  potentials are rational functions, then so are the coefficients of  (\ref{introcurve}). This immediately implies that the branchpoints of (\ref{introcurve}) on the real axis (i.e. the zeroes of the discriminant)  are nowhere dense and hence {\em a priori} the supports of the measures must consist of a finite union of intervals (since they must be compact as shown in Sect. \ref{sect2} in the general setting).

The role of the pseudo--algebraic curve (\ref{introcurve}) is exactly the same as the well--known pseudo--hyperelliptic curve that appears in the one--matrix model \cite{McLaughlinDeiftKriecherbauer, Pastur}: 
%\red
{ in the context of the study of asymptotic properties of multiple orthogonal polynomials it has been pointed out since the fundamental work \cite{Nuttall} that the Cauchy transforms of the extremal measures  solve an algebraic equation}.

%\red
{We also mention the recent work \cite{McLaughlin}, in which the limiting behaviour of Hermitian random matrices with external source is investigated and the presented asymptotic analysis relies on a set of conditions which are shown to be equivalent to the existence of a particular algebraic curve. The methods used in that paper to prove the existence for some special cases are very similar to our approach.}

% \red
{
As it was pointed out by one of the referees, examples of of algebraic curves for special external fields were also obtained in the recent papers \cite{Lapik, Lysov}. } 
\subsection{Connection to a Riemann--Hilbert problem}
The principal motivation to the present paper is the application to the study of biorthogonal (multiply orthogonal) polynomials that arise in the study of the model hinted at by eq. (\ref{12}). In \cite{BertoGekhtSzmig1, BertoGekhtSzmig2} we introduced the biorthogonal polynomials 
\be
\int_{\R_+^2} p_n(x) q_m(y) \frac {{\rm e}^{-N (V_1(x) + V_2(y))}}{x+y} \d x \d y = c_n^2 \delta_{mn}\ ,\ \ p_n(x) = x^n + \dots, \ q_n(y) = y^n+\dots\ .
\ee
In \cite{BertoGekhtSzmig2} it was shown how a natural vector potential problem (for two measures) arises in that context and leads to a three--sheeted spectral curve of the form (\ref{introcurve}). Such problem enters in a natural way in the normalization of the $3\times 3$ Riemann--Hilbert problem considered in \cite{BertoGekhtSzmig1} characterizing those polynomials (and some accessory ones) in the limit $N\to \infty, \ n\to\infty,\ \frac N n\to T>0$. The notation $V_1,V_2$ is meant here to reflect the notation that will be used in Sect. \ref{sect4} and Sect \ref{sec6} (up to a reflection $V_2(y) \mapsto V_2(-y)$, as explained in \cite{BertoGekhtSzmig1,BertoGekhtSzmig2}).

In perspective the more general situation with several measures considered in Sect. \ref{sect3} and Sect. \ref{sect4} will be associated to the polynomials appearing in the study of the random-matrix chain (\ref{chain}) and biorthogonal polynomials for pairings of the form 
\bea
\int_{\R_+^K} p_n(x_1) q_m(x_K) \frac {{\rm e}^{-N \sum_{j=1}^K V_j(x_j)}}{\prod_{j=1}^{K-1}(x_j+ x_{j+1})} \prod_{j=1}^K \d x_j = c_n^2 \delta_{mn}\\
 p_n(x) = x^n + \dots, \ q_n(y) = y^n+\dots
\eea
The details are to appear in forthcoming publications \cite{BertoGekhtSzmig3}.

%%%%%%%%%%%%%%%%%%%%%%%%%%%%%%%%%%%%%%%%%%%%%%%

%% The vector potential problem

%%%%%%%%%%%%%%%%%%%%%%%%%%%%%%%%%%%%%%%%%%%%%%%

\section{The vector potential problem}
\label{sect2}
In this section we consider the vector potential problem which is a slightly generalized form of the weighted energy problem of signed measures (\cite{SaffTotik}, Chapter VIII; \cite{ns}, Chapter 5).

Let $A=(a_{ij})_{i,j=1}^{R}$ be an $R \times R$ real symmetric 
%\red
{{\em positive definite} matrix (in particular it has positive diagonal entries)}, referred to as the 
{\bf interaction matrix}, containing the information on the total charges of the measures and their pair interaction coefficients. 
Suppose $\Sigma_1, \Sigma_2, \dots, \Sigma_R$ is a collection of non-empty, \emph{not necessarily disjoint} closed subsets of $\C$ such that
$\Sigma_k \cap \Sigma_l$ has zero logarithmic capacity whenever $a_{kl} <0$.
Define the functions $h_k \colon \C \to (-\infty,\infty]$ for each $\Sigma_k$ to be
\be
h_k(z) := \ln{\frac{1}{d(z,\Sigma_k)}}, \qquad (z \in \C)
\ee
where $d(\cdot, K)$ is the \emph{distance function} from the closed subset $K$ of the complex plane:
\[d(z, K) := \inf_{t \in K}{|z-t|}.\]
The function $d(z,K)$ is non-negative, uniformly continuous on $\C$ so $h_k(z)$ is upper semi-continuous
and $h_k(z)=\infty$ on $\Sigma_k$.

\begin{definition}
\label{defadm}
A collection of background potentials
\be
V_k \colon \Sigma_k \to (-\infty,\infty], \qquad k=1,2,\dots, R
\ee
is said to be
\emph{admissible with respect to the (positive definite) interaction matrix $A$} if  the following conditions hold:
\renewcommand{\labelenumi}{\bf [A\arabic{enumi}]}
\begin{enumerate}
 \item the potentials $V_k$ are lower semi-continuous on $\Sigma_k$ for all $k$,
 \item the sets $\{z \in \Sigma_k \ \colon \ V_k(z) < \infty\}$ are of positive logarithmic capacity for all $k$,
 \item
 \label{tre}
 the functions
 \be
 H_{jk}(z,t) := \frac {V_j(z)+V_k(t)}{R} + a_{jk} \LN
 \ee
 are uniformly bounded from below, i.e. there exists an $L \in \R$ such that
 \be
 H_{jk}(z,t)\geq L
 \ee
on $\left\{(z,t)\in \Sigma_j\times \Sigma_k\ \colon \ z\not= t\right\}$ for all $j,k=1,\dots, R$.
 Without loss of generality we can assume $L=0$ by adding a common constant to all the potentials so that 
 \be
 H_{jk}(z,t)\geq 0\ .\label{assumptiontre}
 \ee
 We will also assume (again, without loss of generality) that {\bf all the potentials are nonnegative}.
 \item
 \label{quattro}
 There exist  constants $0\leq c<1 $ and $C$  such that (recall that $a_{kk}>0$)
\bea
H_{jk}(z,t)\geq \frac {(1-c)}R(V_j(z)+V_k(t)) -\frac {C}{R^2}. \label{assumptionquattro}
\eea
The constant $C$ can be chosen to be \emph{positive}.
\item
\label{cinque}
The potentials are given such that the functions
\be
\label{growth}
Q_k(z):= \sum_{l:\ a_{kl}<0} \le(\frac 1 R V_l(z)  +  a_{kl} h_l(z)\ri) = \frac {s_k}R V_k(z) +\sum_{l:\ a_{kl}<0}    a_{kl} h_l(z)
\ee
are {\bf bounded from below on }$\Sigma_k$ (here $s_k\leq R-1$ is the number of negative $a_{kl}$'s).\\
Note that in the above sum $k\neq l$ because of the assumption that $a_{kk}>0$.
\end{enumerate}
\end{definition}
\bd
We define the {\bf weighted energy with interaction matrix $A$} of a measure $\vec \mu = [\mu_1,\dots, \mu_R]$ with $\mu_j \in \M_1(\Sigma_j)$ by 
\bea
I_{A,\vec V}(\vec \mu) &\& := \sum_{j,k}^R a_{jk} \iint \LN d\mu_j(z)d\mu_k(t) + 2\sum_{k=1}^R \int V_k(z)d\mu_k(z)\cr
&\& \ =  \sum_{j,k} \iint H_{jk}(z,t) d\mu_j(z) d\mu_k(t),
\eea
where $\M_1(K)$ stands for the set of all Borel probability measures supported on some set $K \subset \C$.
\ed
\br
The assumption {\rm [A\ref{tre}]} is a {\em sufficient} requirement to ensure that the definition of the functional $I_{A, \vec V}(\cdot)$ is well-posed and it is a rather mild assumption on the growth of the potentials near the overlap regions and infinity. Indeed (with $L=0$) 
\be
I_{A,\vec V}(\vec \mu) = \sum_{j,k} \iint H_{jk}(z,t) d\mu_j(z) d\mu_k(t)\geq 0.
\ee
Note also that if a conductor $\Sigma_j$ is unbounded the condition (\ref{assumptionquattro}) implies that 
\be
\frac {c}R V_j(z)\geq   a_{jj}\ln {|z-t_0|} - \frac cR  V_j(t_0) -\frac C {R^2} 
\ee
and hence $V_j$ grows at least like a logarithm. In \cite{SaffTotik} the usual requirement is the stronger one that  $V_j(z)/\ln |z| \to \infty$ as $z\to\infty$.
\er
\br
{\rm [A\ref{quattro}]} is a stronger requirement which will be used for proving tightness (and therefore relative compactness) of a certain subfamily of measures over which $I_{A,\vec V}(\cdot)$ is guaranteed to attain its minimum value.
\er
\br
{\rm [A\ref{cinque}]} is yet stronger and assumes that all potentials have a suitable logarithmic growth near the common boundaries with those condensers carrying an opposite charge. This condition could be relaxed in some settings.
\er

%%%%%%%%%%%%%%%%%%%%%%%%%%%%%%%%%%%%%%%%%%%%%%%

%% Existence and Uniqueness

%%%%%%%%%%%%%%%%%%%%%%%%%%%%%%%%%%%%%%%%%%%%%%%

\section{Existence and uniqueness of the equilibrium measure}
In this section we prove the existence and uniqueness of the equilibrium measure for the vector potential problem described above.
Before stating our main theorem, we recall that a family of measures $\F$ on a metric space $X$ is called {\em tight} if for all $\varepsilon>0$ there exists a compact set $K\subset X$ such that $\mu(X\setminus K)<\varepsilon$ for all measures $\mu \in \F$. The following theorem is a standard result in probability theory:
\bt[Prokhorov's theorem \cite{Prokhorov}]
\label{Prokhorov}
Let $(X,d)$ be a separable metric space and $\M_1(X)$ the set of all Borel probability measures on $X$.
\begin{itemize}
\item If a subset $\F \subset \M_1(X)$ is a {\bf tight family} of measures, then $\F$ is {\em relatively compact} in $\M_1(X)$ in the topology of weak convergence.
\item Conversely, if there exists an equivalent {\em complete} metric $d_0$ on $X$ then every relatively compact subset $\F$ of $\M_1(X)$ is also a tight family. 
\end{itemize}
\et
We will use the following little lemma:
\bl
\label{tightfamily}
Let $F \colon X \to [0, \infty]$ be a non-negative lower semi-continuous function on the locally compact metric space $X$ satisfying
\be
\lim_{x \to \infty} F(x) = \infty,
\ee
i.e. for all $H >0$ there exists a compact set $K \subset X$ such that $F(x) > H$ for all $x \in X\setminus K$.
Then for all $H > \inf{F}$ the family
\be
\F_{H} := \left\{ \mu \in \M_1(X) \ \colon \ \int_{X}Fd\mu < H \right \}
\ee
is a non-empty tight subset of $\M_1(X)$.
\el
{\bf Proof.} $F$ attains its minimum at some point $x_0 \in X$ since $F$ is lower semi-continuous and $\lim_{x \to \infty}F(x)=\infty$ and
therefore the Dirac measure $\delta_{x_0}$ belongs to $\F_{H}$. To prove the tightness of $\F_H$, let $\varepsilon >0$ be given.
Since $F$ goes to infinity ``at the boundary'' of $X$ there exists a compact set $K \subset X$ such that $F(x) > \frac{2H}{\varepsilon}$ for all $x \in X\setminus K$. If $\mu \in \F_{H}$ we have
\be
\mu(X\setminus K) = \int_{X\setminus K} d\mu \leq \frac{\varepsilon}{2H}\int_{X\setminus K} F d\mu \leq \frac{\varepsilon}{2H}\int_{X} F d\mu \leq \frac{\varepsilon}{2H}H = \frac{\varepsilon}{2} < \varepsilon.
\ee
\begin{flushright}
{\bf Q.E.D.}
\end{flushright}

Define
\be
U_{k}^{\vec \mu}(z):= \sum_{k=1}^{R} a_{kl} \int \LN\ d\mu_l(t),
\ee
which is the logarithmic potential (external terms and self-potential together) experienced by the $k$th charge component in the presence of $\vec\mu$ only.
\bt[see \cite{SaffTotik}, Thm. VIII.1.4]
\label{mainthm}
With the admissibility assumptions {\rm [A1]} - {\rm [A5]} above the following statements hold:
\begin{enumerate}
\item The extremal value
\be
\V_{A,\vec V} := \inf_{\vec \mu}{I_{A,\vec V}(\vec \mu)}
\ee
 of the functional $I_{A, \vec V}(\cdot)$ is finite and there exists a unique (vector) measure $\vec \mu^{\star}$ such that $I_{A, \vec V}(\vec \mu) = \V_{A,\vec V}$.

\item The components of $\vec \mu^{\star}$ have finite logarithmic energy and compact support. Moreover, the $V_j$'s and the logarithmic potentials $U_{k}^{\vec \mu^{\star}}$ are bounded on the support of $\mu_k$ for all $k=1,\dots, R$.
\item For $j=1,\dots, R$ the {\em effective potential} 
\be
\varphi_j(z) := U_{j}^{\vec\mu^{\star}}(z) + V_j(z)
\ee
is bounded from below by a constant $F_j$ ({\em Robin's constant}) 
%\red
{on $\Sigma_j$}, 
with the equality holding a.e. on the support of $\mu_j$.
\end{enumerate}
\et
%\red
{\br
The content of 
Thm. \ref{mainthm} is probably neither completely new nor very surprising and the proof is a rather straightforward generalization: the main improvement over the most common literature is the fact that we allow the condensers to overlap even if the corresponding term in the interaction matrix is negative. The assumption on the potentials that they provide a screening effect so that the equilibrium measures will not have support on the overlap region. The theorem will be instrumental in the proof of Thm. \ref{thmalgcurve}, which is the main result of the paper.
\er}
\noindent
{\bf Proof of Theorem \ref{mainthm}.}
First of all, we have to prove that 
\be
\V_{A,\vec V} = \inf_{\vec \mu}{I_{A,\vec V}(\vec \mu)} < \infty
\ee
by showing that there exists a vector measure with finite weighted energy. To this end, let $\vec \eta$ be the $R$-tuple of measures whose $k$th component $\eta_k$ is the equilibrium measure of the standard weighted energy problem (in the sense of \cite{SaffTotik}) with potential $V_k(z)/a_{kk}$ on the conductor $\Sigma_k$ for all $k$.
(The potential $V_k(z)/a_{kk}$ is admissible in the standard sense on $\Sigma_k$ since
\be
\frac{1}{a_{kk}}V_k(z) -\ln|z| \geq \frac{R}{c}\ln {|z-t_0|} - \frac{1}{a_{kk}} V_k(t_0) -\frac{C}{ca_{kk}R} -\ln|z| \to \infty
\ee
as $|z| \to \infty$ for $z \in \Sigma_k$ if $\Sigma_k$ is unbounded.) We know that $\eta_k$ is supported on a compact set of the form
\be
\left\{z \in \Sigma_k \ \colon \ \frac{V_k(z)}{a_{kk}} \leq K_k\right\}
\ee
for some $K_k \in \R$. These sets are mutually disjoint by the growth condition (\ref{growth}) imposed on the potentials. The sum of the ``diagonal'' terms and the potential terms in the energy functional are finite for $\vec \eta$ since this is just a linear combination of the individual weighted energies of the equilibrium measures $\eta_k$. The ``off-diagonal'' terms with positive interaction coefficient $a_{kl}$ are bounded from above because the supports of $\eta_k$ and $\eta_l$ are separated by a positive distance; the terms with negative interaction coefficent are also bounded from above since $\eta_k$ and $\eta_{l}$ are compactly supported. Therefore 
\be
\V_{A,\vec V} \leq I_{A,\vec V}(\vec \eta) < \infty.
\ee

Integrating the inequalities (\ref{assumptionquattro}) it follows that
\bea
I_{A, \vec V}(\vec\mu) &\& = \sum_{j,k=1}^{R} \iint H_{jk}(z,t) d\mu_j(z) d\mu_k(t) \geq (1-c) \sum_{k=1}^R \int V_k(z)d\mu_k(z) - C.
\eea
We then study the minimization problem over the following set of probability measures:
\be
\label{condition}
\F:= \le\{ \vec \mu  \ \colon\ \sum_{k=1}^{R} \int V_k(z) d\mu_k(z)\leq \frac 1{(1-c)} (\mathcal V_{A,\vec V} + C+1)  \ri\} \subset \M_1(\Sigma_1)\times  \dots \times \M_1( \Sigma_R)\ .
\ee
 The extremal measure(s) are all contained in $\F$ since for a vector measure $\vec \lambda\not \in \F$ we have
\be
I_{A,\vec V}(\vec \lambda) \geq (1-c) \sum_{k=1}^R \int V_k(z)d\lambda_k(z) - C \geq \V_{A,\vec V }+1.
\ee
The function $\sum_k{V_k(z)}$ is non-negative, lower semi-continuous and goes to infinity as $|z| \to \infty$, and moreover
\be
\frac{R}{(1-c)} (\mathcal V_{A,\vec V} + C+1) > 0,
\ee
hence, by Lemma \ref{tightfamily}, all projections of  $\F$ to the individual factors is a non-empty tight family of measures.
Using Prokhorov's Theorem \ref{Prokhorov}  we know that there exists a measure $\vec \mu^{\star}$ minimizing $I_{A,\vec V}(\cdot)$ such that $\frac{1}{R}\sum_{k=1}^{R}\mu_\star \in \F$. The existence of the (vector) equilibrium measure is therefore established.

Note that now statement $(2)$ follows immediately: indeed from the condition \ref{tre} that $H_{j,k}\geq 0$ (and also $V_j\geq 0$)  it follows that 
\bea
\mathcal V_{A,\vec V} & = & a_{11}\iint \LN d\mu^{\star}_1(z) d\mu^{\star}_1(t) + \frac{2}{R} \int V_1(z) d\mu^{\star}_1(z) \cr 
&& + \sum_{(j, k)\neq (1,1)} \iint  H_{jk}(z,t)d\mu^{\star}_j(z) d\mu^{\star}_k(t) \cr
&\geq &   a_{11}\iint \LN d\mu^{\star}_1(z) d\mu^{\star}_1(t).
\eea
Thus the logarithmic energy of $\mu^{\star}_1$ is bounded above by $\mathcal V_{A,\vec V}/a_{11}$.
Repeating the argument for all $\mu^{\star}_j$'s we have that  all the logarithmic energies of the $\mu^{\star}_j$'s are bounded above. 

On the other hand, these log-energies are also bounded below using (\ref{assumptionquattro}) with $j=k$:
\be
a_{jj}\iint\LN d\mu^{\star}_j(z)d\mu^{\star}_j(t)  \geq -\frac {2c}R \int V_j(z)d\mu^{\star}_j(z)  - \frac C{R^2}
\ee
(boundedness from below follows since $ \int V_j(z)d\mu_j(z)$ is bounded above and appears with a negative coefficient in the formula).

Now, using the fact that the quantities $H_{jk}(z,t)$
are nonnegative due to (\ref{assumptiontre})  and condition (\ref{condition}) it follows that 
\be
\varphi_j(z) = V_j(z) + \sum_{k\neq j}a_{jk}  \int \LN d\mu^{\star}_{k}(t) 
\ee
is finite wherever $V_j(z)$ is. Using condition {\rm [A\ref{cinque}]} it also follows that it is lower semicontinuous,  bounded from below on $\Sigma_j$ and hence admissible in the usual sense of minimizations of single measures \cite{SaffTotik}. 
We also claim that $\varphi_j$ grows to infinity near all the contacts between $\Sigma_j$ and any $\Sigma_k$ for which $a_{jk}<0$.  Suppose $z_0\in \Sigma_j\cap \Sigma_k $ (with $a_{jk}<0$); then on a  compact neighborhood $K$  of $z_0$ we have 
\be
\varphi_j(z) \geq V_j(z) + \sum_{k\neq j \atop  a_{jk}<0} a_{jk} h_k(z) + M_K 
\ee
for some finite constant $M_K$ (which --of course-- depends on $K$).
From (\ref{cinque}) then 
\be
 V_j(z) + \sum_{k\neq j \atop  a_{jk}<0} a_{jk} h_k(z) + M_K  \geq  \frac{R-s_j} R V_j(z) + \wt M_K
\ee
where $s_j<R$ is the number of negative  $a_{jk}$ ($j\neq k$). Since $V_j(z)$ tends to infinity at the contact points (from the same condition {\rm [A\ref{cinque}]}) then so must be for $\varphi_j$.

Note also  that 
\be
\mathcal{V}_{A,\vec V} = \sum_j I_{\Sigma_j, \varphi_j}\le(\mu_{j,\star}\ri), 
\ee
and hence (as in \cite{SaffTotik}) each single $\mu_{j,\star}$ is the minimizer of the single variational problem on $\Sigma_j$ under the effective potential $\varphi_j$. From the standard results it follows that the support of $\mu^{\star}_{j}$ is contained in the set where $\varphi_j$ is bounded, which, due to our assumptions, are all compact and at finite nonzero distance from the common overlaps. This proves that the components of $\vec \mu^{\star}$ are actually compactly supported.

Uniqueness as well as the remaining properties are established essentially in the same way as in \cite{SaffTotik}, Thm. 1 Chap. VIII 
%\red
{using the positive definiteness of the interaction matrix $A$, which guarantees the convexity of the functional.}
\begin{flushright}
{\bf Q.E.D.}
\end{flushright}

\section{The special case}
\label{sect3}

We now specialize the above setting to the following collection of $R$ conductors:
\be
\Sigma_j := (-1)^{j-1} [0,\infty) \qquad (j=1,2,\dots, R),
\ee
and interaction matrix
\be
A := \left[
\begin{array}{ccccc}
2q_1^2 & -q_1 q_2 & 0 & \ldots & 0 \\
-q_1 q_2 &2q_2^2 & - q_2 q_3  & \ldots & 0 \\
0 & -q_2 q_3 & 2q_3^2 & \ldots & 0 \\
\vdots & \vdots & \vdots & \ddots & \vdots \\
0 & 0 & 0 & \dots & 2q_R^2 
\end{array}
\right].
\ee

Under the assumptions on the growth of the potentials $V_j(x)$ near the only common boundary point $x=0$, Thm. \ref{mainthm} guarantees the existence of a unique vector minimizer. 

We now investigate the regularity properties under the rather comfortable assumption that the potentials $V_j$ are \emph{real analytic} on $\Sigma_j\setminus \{0\}$ for all $j$; 
%\red
{this is in addition to the host of assumptions specified in Def. \ref{defadm}.}

In order to simplify slightly some algebraic manipulations to come we re-define the problem by rescaling the component of the vector of probability measures $\mu_j\mapsto q_j\mu_j$ so that now the interaction matrix becomes the simpler
\be
A:= \left[
\begin{array}{ccccc}
2 & -1 & 0 & \ldots & 0 \\
-1 & 2 & - 1  & \ldots & 0 \\
0 & -1 & 2 & \ldots & 0 \\
\vdots & \vdots & \vdots & \ddots & \vdots \\
0 & 0 & 0 & \dots &2
\end{array}
\right].
\ee

The electrostatic energy can be rewritten as
\bea
I_{A,\vec V}(\vec \mu) &=&  2\sum_{j=1}^R\iint \ln \frac 1{|x-y|} d\mu_j(x)d\mu_j(y) -  \sum_{j=1}^{R-1}\iint \ln \frac 1{|x-y|} d\mu_j(x)d\mu_{j+1}(y)\\
&& + 2\sum_{j=1}^R \int V_j(x) d\mu_j(x).
\eea

As explained in the previous section, the above minimization problem has the interesting property that the same equilibrium measure is achieved by minimizing only one component  of it in the mean field of the neighbors 
%\red
{and, moreover, the supports of the minimizers satisfy 
\be
\supp(\rho_j) \cap \supp(\rho_{j+1}) = \emptyset\ .\label{nointersect}
\ee}

\bc
\label{Coreffect}
Let $\vec \mu$ be the vector equilibrium measure for the above problem. For any $1\leq k\leq R$ we have that 
\be
I_{\wh V_{k}}(\mu):= \int_{\Sigma_k}\int_{\Sigma_k}\ln\frac{1}{|z-t|}d\mu(z) d\mu(t) + 2\int_{\Sigma_k} \wh V_{k}(z) \d\mu(z)
\ee
is minimized by the same $\mu_k$, 
where the effective potentials $\wh V_k $ are 
\bea
\wh V_1(z) &:=&\frac 1 2  V_1(z)  - \frac 1 2 \int_{\Sigma_{2}} \ln\frac{1}{|z-t|} d\mu_{2}(t)\\
\wh V_k(z) &:=&\frac 1 2  V_k(z)  - \frac 1 2 \int_{\Sigma_{k+1}} \ln\frac{1}{|z-t|} d\mu_{k+1}(t) -\frac 1 2 \int_{\Sigma_{k-1}} \ln\frac{1}{|z-t|} d\mu_{k-1}(t)\\
\wh V_R(z) &:=&\frac 1 2  V_R(z)  - \frac 1 2 \int_{\Sigma_{R-1}} \ln\frac{1}{|z-t|} d\mu_{R-1}(t).
\eea
\ec
Note that the effective potential differs from the original potential by {\bf harmonic} potentials because  the supports of $\mu_{k\pm 1}$ are disjoint from the support of $\mu_k$.

%We now {\bf specialize} the abve setting to $\Sigma_j$ being closed intervals of $\R$ with otherwise the same properties as above.
We recall the following theorem:
\bt[Thm. 1.34 in \cite{McLaughlinDeiftKriecherbauer}]
\label{thmMDK}
If the external potential belongs to the class $\mathcal C^k$, $k\geq 3$ then the equilibrium measure is absolutely continuous and its density is H\"older continuous of order $\frac 1 2$.
\et

Combining Cor. \ref{Coreffect} with Thm. \ref{thmMDK} we have that  the solution of our equilibrium problem consists of equilibrium measures which are absolutely continuous with respect to the Lebesgue measure with densities $\rho_j$ at least H\"older--$\frac 12$ continuous as long as the external potentials are at least $\mathcal C^3$.
Moreover the supports of these equilibrium measures have a {\em finite positive distance} from the origin.

 Our next goal is to prove that the supports of the $\rho_j$'s consist of a finite union of disjoint compact intervals. For that we need a pseudo--algebraic curve given in the next section.

\section{Spectral curve}
\label{sect4}
Since the equilibrium measures have a smooth density we can now proceed with some manipulations using the variational equations.

For the remainder of the paper we will make the following additional {\bf assumption} 
%\red
{(besides those in Def. \ref{defadm})} on the nature of the potentials $V_j$:

{\bf Assumption}: \emph{the derivative of the potential $V'_j$ is the restriction to $\Sigma_j^o:= (-1)^{j-1} (0,\infty)$ of a real analytic function defined in a neighborhood of  the real axis possessing at most isolated {\em polar} singularities on $\R\setminus \Sigma_j$}.

For a function $f$ analytic on $\C\setminus \Gamma$, where $\Gamma$ is an oriented smooth curve, we denote
\be
\mathcal S (f)(x) := f_+(x)  + f_-(x)\ ,\ \ \ 
\Delta (f)(x) := f_+(x) - f_-(x)\ ,\ \ x\in \Gamma.
\ee
where the subscripts denote the boundary values.
%\red
{We remind the reader that under our assumptions, the equilibrium measures satisfy eq. (\ref{nointersect}). }
\bd
For the solution $\vec \rho$ of the variational problem, we define the {\bf resolvents} as the expressions
\be
W_j(z):=  \int_{\Sigma_j} \frac {\rho_j(x)\d x}{z-x}\ ,\qquad z\in \C\setminus \supp(\rho_j).
\ee
\ed

The variational equations imply the following identities for $j=1,\dots, R$:
\bea
\mathcal S(W_j)(x)&\&  =  V'_j(x) + W_{j+1} + W_{j-1}\cr
\Delta(W_j)(x) &\& =- 2i\pi \rho_j(x) ,\qquad x\in \supp(\rho_j)
\eea
where we have convened that $W_0\equiv W_{R+1} \equiv 0$.
Note that, under our assumptions for the growth of the potentials $V_j$ 
%\red
{at the contact points between conductors (in this case the origin),} the support of $\rho_j$ is disjoint from the supports of $\rho_{j\pm 1}$ and hence the resolvents on the rhs of the above equation are continuous on $\supp(\rho_j)$.

The following manipulations are purely algebraic: we first introduce the new vector of functions
\be
\le[
\begin{array}{c}
Y_1\\\vdots\\ Y_R
\end{array}
\ri]^t :=  \le[\begin{array}{cccc}
-1 & &&\\
&1&&\\
&&\ddots&\\
&&&(-1)^{R}
\end{array}\ri]\le\{A^{-1}\le[\begin{array}{c}
V_1'\\
\vdots\\
V_R'
\end{array}\ri]  + \le[\begin{array}{c}
W_1\\
\vdots\\
W_R
\end{array}\ri] \ri\}
\label{shiftresolvents}
\ee
Trivial linear algebra implies then the following relations for the newly defined functions $Y_j$:
\be
\begin{array}{rclrclc}
\mathcal S(Y_1) &=& -Y_2 & \Delta(Y_1) &=& 2i\pi \rho_1 & \hbox { on } \supp(\rho_1)\\
\mathcal S(Y_2) &=& -Y_1-Y_3 & \Delta(Y_2) &=&-2i\pi \rho_2 & \hbox { on } \supp(\rho_2)\\
\mathcal S(Y_3) &=& -Y_2-Y_4 & \Delta(Y_3) &=& 2i\pi \rho_3 & \hbox { on } \supp(\rho_3)\\
&\vdots &&& \vdots &&\vdots\\
\mathcal S(Y_{R-1}) &=& -Y_{R-2}-Y_R & \Delta(Y_{R-1}) &=& (-1)^{R}2i\pi \rho_{R-1} & \hbox { on } \supp(\rho_{R-1})\\
\mathcal S(Y_R) &=& -Y_{R-1} & \Delta(Y_R) &=& (-1)^{R+1}2i\pi \rho_R & \hbox { on } \supp(\rho_R).
\end{array}
\ee
The above relation should be understood at all points that do not coincide with some of the isolated singularities of some potential $V_j$ (points of which type there are only finitely many within any compact set).

Define then the functions 
\be
Z_0:= Y_1\ ,\ \ Z_1 := -Y_1-Y_2\ ,\ \ Z_2:= Y_2+Y_3\ ,\dots, \ Z_{R-1}  = (-1)^{R-1}(Y_{R-1}+Y_R)\ ,\ \ Z_R := (-1)^{R} Y_R.
\ee
Then 
\bp
\label{propnocuts}
All symmetric polynomials of $\{Z_j\}_{0\leq j\leq R}$ are {\em real analytic} in the common domain of analyticity of the potentials, namely they have no discontinuities on the supports of the measures $\rho_j$.
\ep
{\bf Proof.}
A direct algebraic computation using the boundary values of the $\{Y_j\}$ functions gives the following boundary values of the functions $Z_j$:
\bea
2Z_{0_\pm } &\& = -Y_2 \pm 2i\pi \rho_1\\
2Z_{1_\pm}&\& = \le\{
\begin{array}{ll}
-Y_2 \mp 2i\pi \rho_1 = 2Z_{0\mp} & \hbox { on } \supp(\rho_1)\\
-Y_1+Y_3 \pm 2i\pi \rho_2 &\hbox { on } \supp(\rho_2) 
\end{array}
\ri.\\
2Z_{2_\pm}&\& = \le\{
\begin{array}{ll}
-Y_1+Y_3 \mp 2i\pi \rho_2 = 2Z_{1\mp} & \hbox { on } \supp(\rho_2)\\
Y_2-Y_4 \pm 2i\pi \rho_3 &\hbox { on } \supp(\rho_3)
\end{array}
\ri.\\
&\vdots &\\
2Z_{(R-1)_\pm}&\& = \le\{
\begin{array}{ll}
(-1)^{R-1}(-Y_{R-2}+Y_{R}) \mp 2i\pi \rho_{R-1} = 2Z_{(R-2)_\mp} & \hbox { on } \supp(\rho_{R-1})\\
(-1)^{R-1} Y_{R-1} \pm 2i\pi \rho_R &\hbox { on } \supp(\rho_R) 
\end{array}
\ri.\\
2Z_{R_\pm} &\&  = (-1)^{R-1} Y_{R-1} \mp 2i\pi \rho_{R} = 2Z_{(R-1)_\mp} \qquad \hbox{ on } \supp(\rho_{R})
\eea
Consider a symmetric polynomial $P_K:= 2^K \le({Z_0}^K + \dots + {Z_R}^K\ri)$ and its boundary values on, say, $\supp(\rho_1)$; we see above that $Z_{0_\pm} = Z_{1_\mp}$ and hence $Z_{0}^K + Z_{1}^K$ has no jump there. The support of $\rho_2$ has no intersection with $\Sigma_1$ 
%\red
{and $\supp(\rho_1)$ (see  (\ref{nointersect})) due to our assumptions},  and hence $Z_2$ may have a jump on $\supp(\rho_1)$ only if the support of $\rho_3$ has some intersection with it. In that case anyway $Z_{2_\pm} = Z_{3_\mp}$ and hence also $Z_{2}^K+Z_{3}^K$ has no jump on $\supp(\rho_3)\cap \supp(\rho_1)$.

In general on $\supp(\rho_k)\cap \supp(\rho_1)$ we have $Z_{k_\pm} = Z_{k_\mp}$ and so the same argument apply. In short one can thus check that all the jumps that may {\em a priori} occur in fact cancel out in a similar way.

Repeating the argument for all the other $\supp(\rho_j)$ instead of $\supp(\rho_1)$ proves that the expression has no jump on any of the supports, and since {\em a priori} it can have jumps only there, then it has  no jumps at all. 
%\red
{Invoking Morera's theorem, we see that the symmetric polynomials of the $Z_k$'s can be extended analytically across the supports of the $\rho_j$'s.}

Finally, the statement that the symmetric polynomials are real analytic follows from the following reasoning: the $Z_j$'s are linear expressions in the $W_j$'s and the potentials. In particular they are analytic off the real axis (where all the $W_j$'s are) and in the common domain of analyticity of the potentials. The same then applies to the symmetric polynomials in the $Z_j$'s. 
Finally, on an open interval in $\R$, as long as it is outside of all the supports of the vector measure, the $Z_j$ are all real analytic functions since $W_j$'s are. This concludes the proof. 
\begin{flushright}
{\bf Q.E.D.}
\end{flushright}
\par \vskip 5pt

A consequence of this proposition is that 

\bt
\label{thmalgcurve}
The functions $Z_k$ are solution of a pseudo--algebraic equation of the form 
\be
z^{R+1} + C_2(x) z^{R-1} + \dots + C_{R+1}(x) = 0\label{algcurve}
\ee
where $C_j(x) := (-1)^j \sum_{\ell_1,\dots,\ell_j} Z_{\ell_1} \cdots Z_{\ell_j}$ are (real) analytic functions on the common domain of analyticity of the potentials.
\et
%\red
{\br
This result is the direct analogue of the results about the existence of the spectral curve for the one-matrix model \cite{Pastur} which was established on a rigorous ground in \cite{McLaughlinDeiftKriecherbauer}. In a different context of matrix models with external source Thm. \ref{thmalgcurve} is conceptually similar to the result in \cite{McLaughlin}.
\er}
{\bf  Proof of Thm. \ref{thmalgcurve}}
We set 
\be
E(z,x):= \prod_{j=0}^{R} (z-Z_j(x))\ ,
\ee
and expand the polynomial in $z$.
Clearly we have $Z_0 + Z_1 + \dots + Z_{R} =0$ and hence the coefficient $C_1$ vanishes identically. The other coefficients are polynomials in the elementary symmetric functions already shown to be real analytic and hence sharing the same property.
\begin{flushright}
{\bf Q.E.D.}
\end{flushright}
\par \vskip 5pt
\begin{wrapfigure}{r}{0.5\textwidth}
\resizebox{0.45 \textwidth}{!}{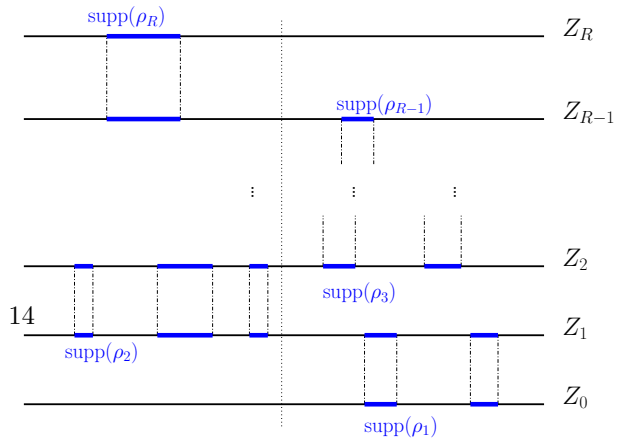}
\caption{The Hurwitz diagram of the spectral curve.}
\label{figtwo}
\end{wrapfigure}
\bc
The densities $\rho_j$ are supported on a {\bf finite union of compact intervals}.
Moreover the supports of $\rho_j$ and $\rho_{j\pm 1}$ are disjoint.
\ec
{\bf Proof}.

The supports of the measures are in correspondence with the jumps of the algebraic solutions of $E(z,x)=0$; in particular the set of endpoints of the supports must be a subset of the zeroes or poles of the discriminant that belong to $\R$. Since the only singularities that these may have come from those of the derivatives of the potentials $V'_j(x)$ on the real axis, and these have been assumed to be meromorphic on $\R$ and be otherwise real analytic, then the discriminant of the pseudo--algebraic equation cannot have infinitely many zeroes on a compact set. We also know that the measures $\rho_j$ are compactly supported a priori and hence there can be only finitely many intervals of support.
\begin{flushright}
{\bf Q.E.D.}
\end{flushright}
\par \vskip 4pt
Putting together Prop. \ref{propnocuts} and Thm. \ref{thmalgcurve} we can rephrase the properties of the functions $Z_j(x)$ by saying that they are the $R+1$ branches of the  polynomial equation (\ref{algcurve}), thus defining an $(R+1)$--fold covering of (a neighborhood of) the real axis. The neighborhood is the maximal common neighborhood of joint analyticity of the potentials $V_j(x)$. 
The various {\em sheets} defined by the functions $Z_j(z)$ are glued toghether along the supports of the equilibrium measures $\rho_j$ in  a ``chain'' of sheets as the {\em Hurwitz} diagram in Fig. \ref{figtwo} shows.

\br
%\red
{
In \cite{Aptekarev} a similar problem was considered in the context of multiple orthogonality for Nikishin systems on conductors without  intersection and with fixed weights: this corresponds to the case of a minimization problem without external fields. It was shown that an algebraic curve similarly arises; in the formulation of \cite{Aptekarev}  the algebraic curve involves, rather than the resolvents, their exponentiated antiderivatives $\Psi_j$'s, namely 
\be
W_j= \frac {\d} {\d x} \ln \Psi_j(x)
\ee
and a mixture of algebraic geometry and geometric function theory was used to investigate their properties. In particular the functions $\Psi_j$ figured in an algebraic equation (see eq. (2.1) in \cite{Aptekarev}) as the various determinations of a polynomial relation
\be
\Psi^{R+1}  + r_{_1}(x) \Psi^R + \dots + r_{_R}(x)\Psi + r_{_{R+1}}(x) = 0\ ,\ \ r_j \in \C[x]
\ee
with the discriminant (w.r.t. $\Psi$) vanishing at the endpoints of the supports for the measures of the corresponding Nikishin problem.
Along similar lines, examples of curves of algebraic type for Nikishin systems with special choices of external fields were recently obtained in \cite{Lapik}.
}
\er

\section{An explicit example}
\label{sec6}
We consider the case with $R=2$ and  the two potentials are the same  $V_1(x)=V_2(-x)$ and are of the simplest possible form that satisfies our requirements
\bea
V_1(x) = bx - a\ln x\ ,\qquad x>0\ ;\ \ 
V_2(x) = -b x -a\ln (-x)\ ,\qquad x<0
\eea
where both $a,b>0$.

Quite clearly we can rescale the axis  and set $b=1$ without loss of generality.

Using the expressions for the coefficients of the spectral curve (Thm. \ref{algcurve}) in terms of the potentials $V_1 = V $ and $V_2 =V^\star = V(-x)$ we have
\bea
&& E(z,x) = z^3 - R(x)z - D(x)=0
\eea
where, on account of the fact that the derivative of the potentials have a simple pole at $x=0$, the coefficients $R(x), D(x)$ have at most a double pole there.
From the relationship between the three branches of $Z$ and the resolvents $W_1,W_2$ (eq. \ref{shiftresolvents}) we have
\bea
Z^{(0)}(x)  &\& = -W_1 -\frac a x + \frac 1 3\\
Z^{(2)}(x)&\& = W_2 + \frac a x + \frac 1 3\\
Z^{(1)}(x)&\&  = -Z^{(0)}(x) - Z^{(2)}(x)  = W_1(x)-W_2(x) + \frac {2a}x 
\eea
and hence the general forms that we can expect for the coefficients of the algebraic curve are
\bea
&& R(x) =  \frac{a^2}{x^2} +\frac 1 3  + \frac C x \cr
&& D(x) = \frac {2a^2}{3x^2} - \frac 2 {27} + \frac {A}{x^2} + \frac {B}x
\eea
where the constants $A,B,C$ have yet to be determined.

The spectral curve $z^3-Rz-D=0$ has in general $5$ finite branchpoints (which is incompatible with the requirements of compactness of the support of the measures) and requiring that there are $\leq 4$ branchpoints and symmetrically placed around the origin(by looking at the discriminant of the equation) imposes that $B=C=0$.

The ensuing spectral curve is 
\be
z^3 - \le(\frac 1 3 + \frac {a^2}{x^2} \ri)z -\le( \frac {2a^2+3A}{3x^2} - \frac 2 {27} \ri)=0
\ee
 and a suitable  rational uniformization of this curve is 
 \bea
 X &\&= \frac{\sqrt{a^2+A}}\lambda   - \frac {A}{2\sqrt{a^2+A}} \le(\frac 1{\lambda+1} + \frac 1{\lambda-1} \ri)\\
Z&\& = -\frac {3A+2a^2}{3a^2} -  \frac{A(a^2+A)}{\le(\lambda^2 - (1+A/a^2)\ri)a^4}
 \eea
The three points above $x=\infty$ are $\lambda=\pm 1, 0$ and $Z$ is regular there.

We see that the condition that the measures $\rho_1,\rho_2$ have unit mass requires that 
\be
\res{x=\infty} Z^{(0)} \d x =  1 + a\ ,\qquad 
\res{x=\infty} Z^{(2)} \d x = -1-a\ .
\ee
We need only to decide which point $\lambda=\pm 1,0$ correspond to the three points over infinity. 
But this is achieved by inspection of the behavior of $Y(\lambda)$ and $X(\lambda)$ near the three points $\lambda=0,1,-1$.
0.

By this inspection we have
\bea
\lambda=1 \ \leftrightarrow \ \infty_1\\
\lambda=-1\ \leftrightarrow\ \infty_2\\
\lambda=0\ \leftrightarrow \ \infty_0\ .
\eea
Computing the residues of $Z\d x = Z X' \d \lambda$ at these points we have 
\bea
&& \res{x=\infty}Z^{(0)} \d x =  \sqrt{a^2+A} = 1 + a\\
&& \res{x=\infty}Z^{(2)} \d x =  -\sqrt{a^2+A} = -1 - a
\eea
which imply that $
A = 2a+1$.

Collecting the above, we have found that 
\bea
 X &\&= \frac{a+1}\lambda   - \frac {2a+1}{2a+2} \le(\frac 1{\lambda+1} + \frac 1{\lambda-1} \ri)\\
Z&\& = -\frac {2a^2 +6a+3}{3a^2} -  \frac{(2a+1)(a+1)}{\le(\lambda^2 - ((a+1)^2/a^2)\ri)a^4}
 \eea
and the algebraic equation for $z = Z(\lambda)$ in terms of $x=X(\lambda)$ becomes
 
\bea
z^3 - \le(\frac 1 3 + \frac {a^2}{x^2} \ri)z -\le( \frac {2a^2+6a+6}{3x^2} - \frac 2 {27} \ri)=0
\eea

\begin{figure}
\resizebox{4cm}{!}{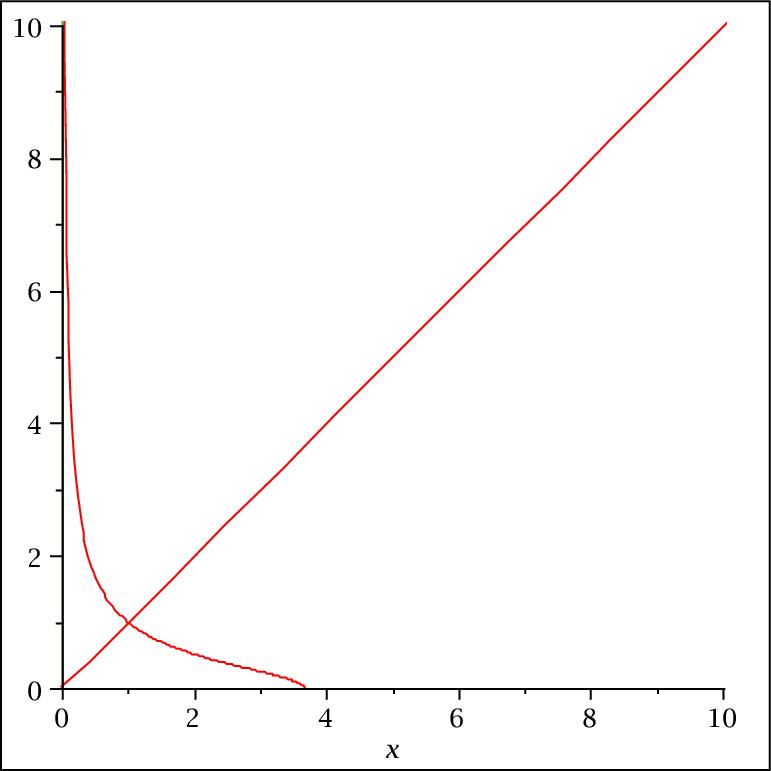}\resizebox{4cm}{!}{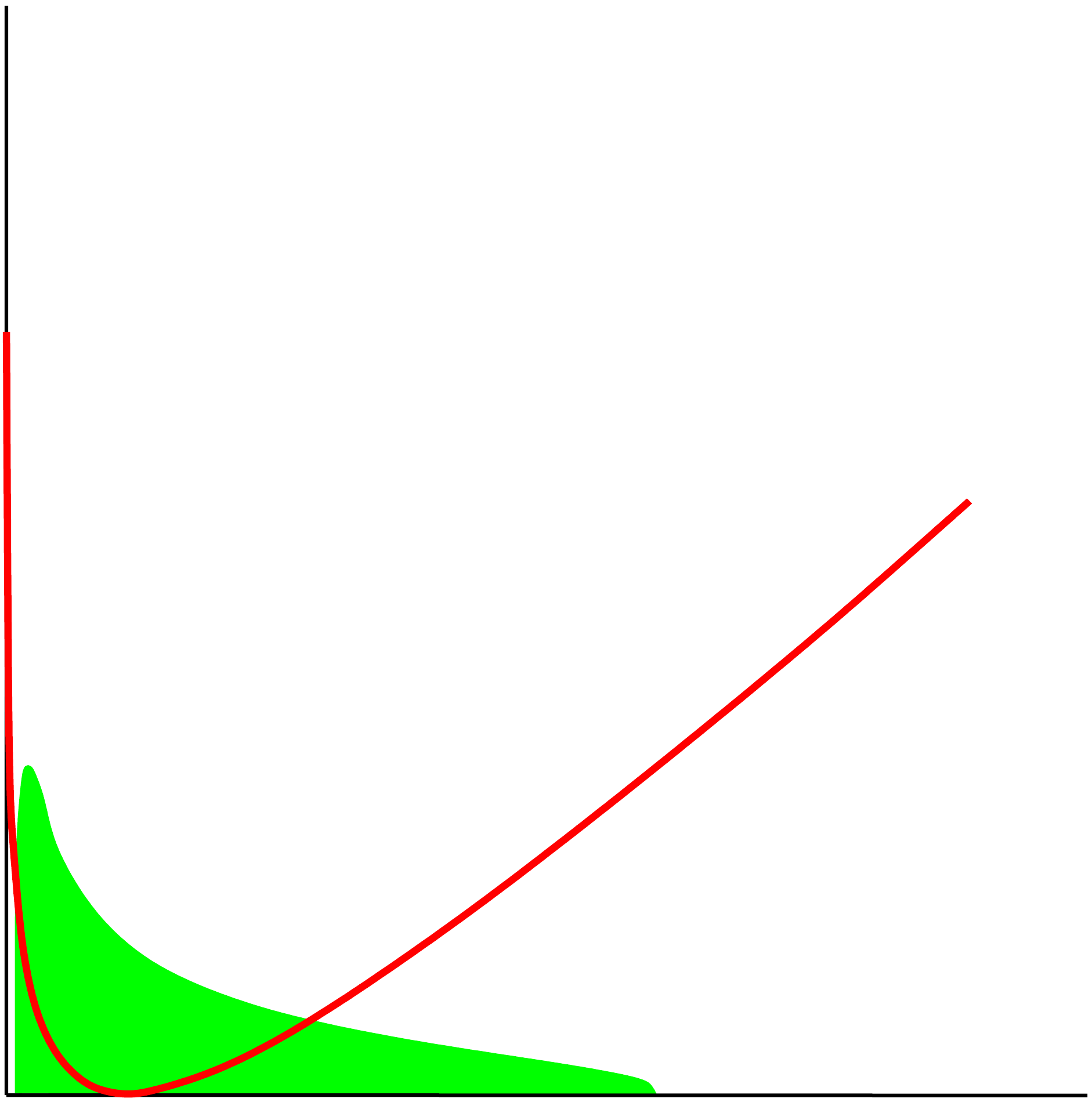}\resizebox{4cm}{!}{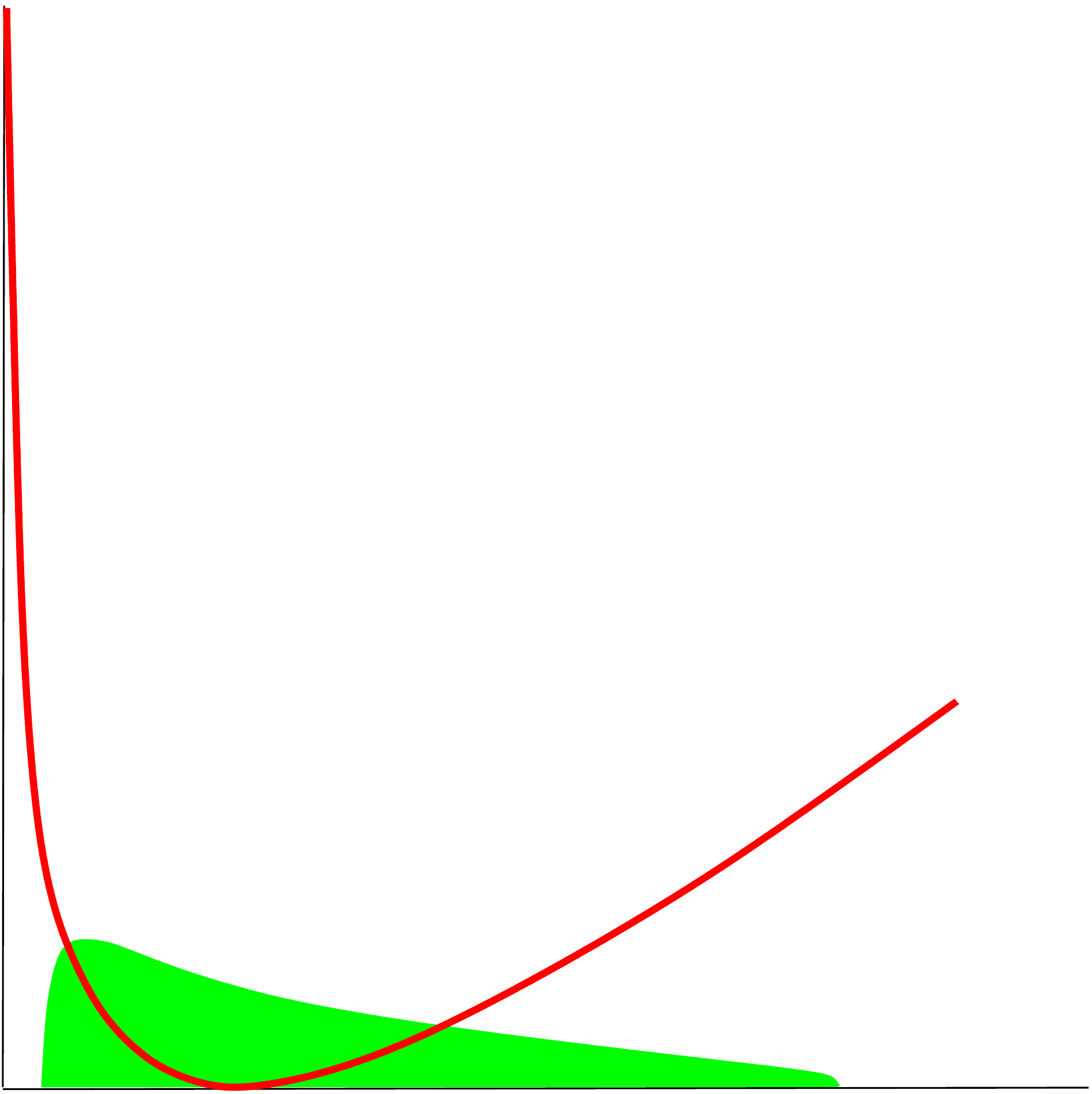}\resizebox{4cm}{!}{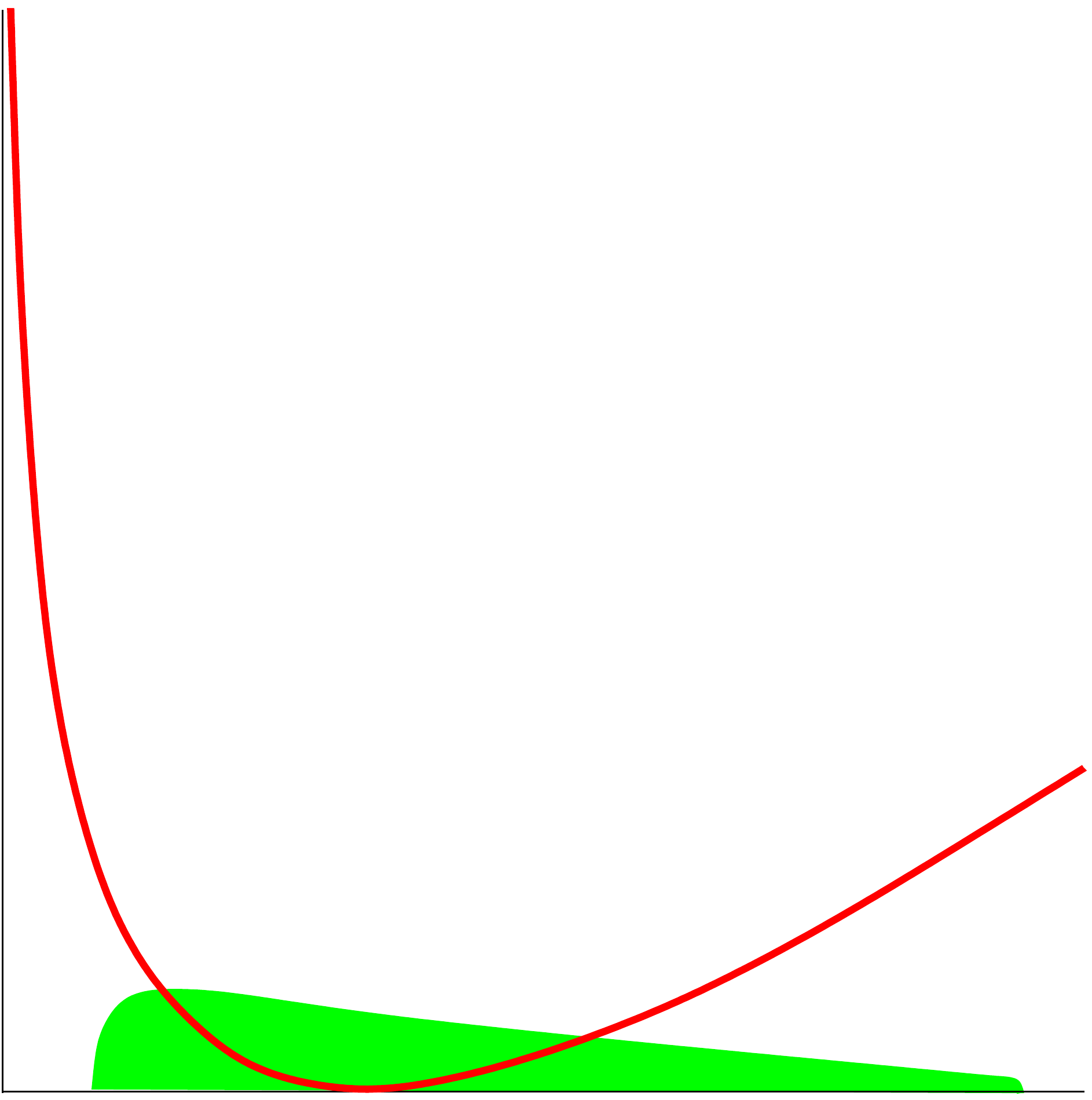}
\caption{Some examples for the equilibrium measure for the example worked out in the text, and $a=0,1,2,3$ respectively from left to right. In red is the graph of the potential $V_1$. The symmetry implies that the other equilibrium measure is simply the reflection of this around the ordinate axis. The units for the axes are the same in all cases. The growth of the density at $x=0$ for $a=0$ is $\mathcal O(x^{-2/3})$. Near the other edges the vanishing is of the form $\mathcal O( (x-\alpha)^{\frac 12 })$. }
\end{figure}

It is possible to write explicitly the expressions of the branchpoints in terms of $a$ but it is not very interesting per se, except to discuss their behaviors in different regimes of $a$;
we find that for $a>0$ there are four symmetric branchpoints on the real axis and the inmost ones tend to zero as $a\to 0$, whereas they all tend to infinity as $a\to\infty$ according to $\pm (a \pm 2 \sqrt{a}) + \mathcal O(1)$.

\begin{wrapfigure}{r}{0.5\textwidth}
\resizebox{0.45\textwidth}{!}{
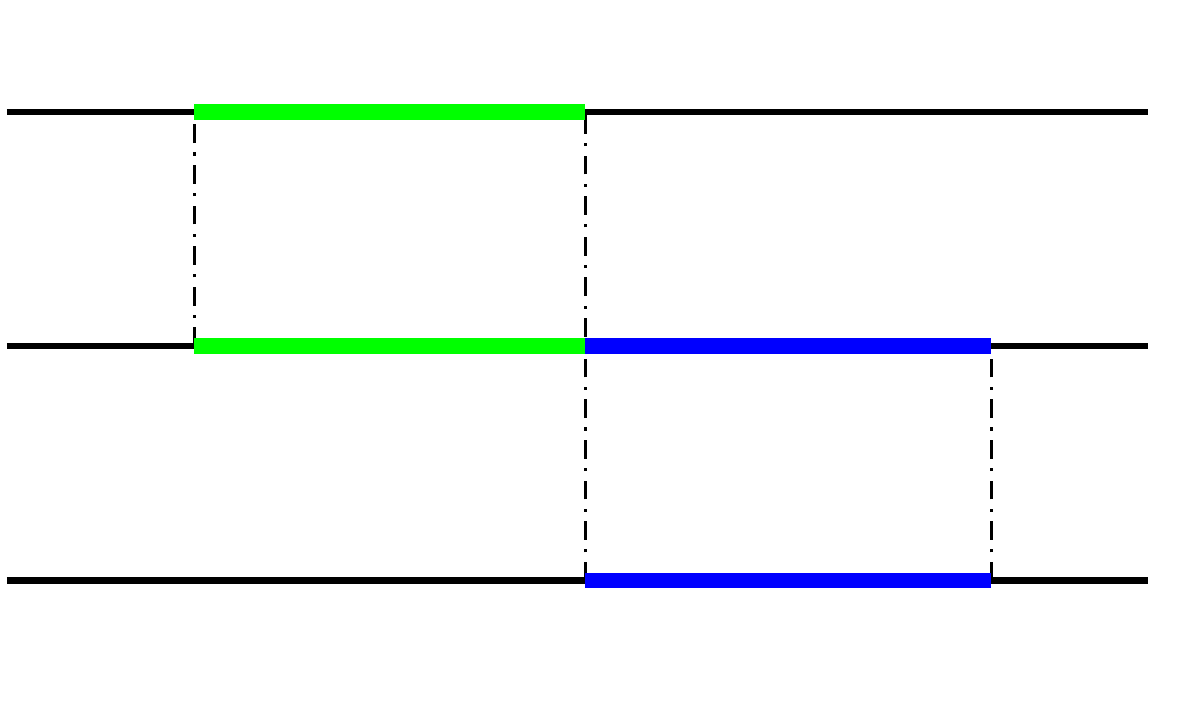}
\end{wrapfigure}

It is interesting to note that for $a=0$ our general theorem does not apply: the potentials are {\em finite} on the common boundary of the condensers and hence cannot prevent accumulation of charge there. However the algebraic solution we have obtained is perfectly well--defined for $a=0$
giving the algebraic relation
\be
z^3 - \frac  z {3} - \frac {2}{x^2} + \frac 2 {27}=0
\ee
A short exercise using Cardano's formul\ae\ shows that the origin is a branchpoint of order $3$ and thus corresponding to the Hurwitz diagram on the side.

The behavior of the equilibrium densities $\rho_j$ near the origin is (expectedly) $x^{-\frac 23}$.
\section{Concluding remarks}
We point out a few shortcomings and interesting open questions about the above problem.

The first problem would be to relax the growth condition of the potentials near common points of boundaries, if not in the general case at least in the specific example given in the second half of the paper, where we consider conductors being subsets of the real axis. 

The importance of this setup is in relation to the asymptotic analysis of certain biorthogonal polynomials studied elsewhere \cite{BertoGekhtSzmig1} and their relationship with a random  multi--matrix model \cite{BertoGekhtSzmig2}.

In that setting, having {\em bounded} potentials near the origin $0\in \R$ would allow the occurrence of new universality classes where new parametrices for the corresponding $3\times 3$ (in the simplest case) Riemann--Hilbert problem would have to be constructed.

Based on heuristic considerations involving the analysis of the spectral curve of said RH problems, the density of eigenvalues should have a behavior of type $x^{-\frac 23}$ near the origin (to be compared with $x^{-\frac 12}$ for the usual hard--edge in the Hermitian matrix model). Generalization involving chain matrix model would allow arbitrary $-\frac pq$ behavior, $p<q$. However, for all these analyses to take place the corresponding equilibrium problem should be analyzed from the point of view of potential theory, allowing bounded potentials near the point of contact.
\section{Acknowledgement}
This research was partially supported by NSERC Discovery grant ``Exact and asymptotic methods in Random Matrtix Theory and Integrable Systems'' and the FQRNT ``Projet de recherche en \'equipe: Th\'eorie spectrale des matrices al\'eatoires''.
\bibliographystyle{unsrt}
\bibliography{Biblio}
\end{document}

%% file: Hurwitzdiagram.pdf_t
\begin{picture}(0,0)%
\includegraphics{Hurwitzdiagram.pdf}%
\end{picture}%
\setlength{\unitlength}{3947sp}%
\begingroup\makeatletter\ifx\SetFigFont\undefined%
\gdef\SetFigFont#1#2#3#4#5{%
  \reset@font\fontsize{#1}{#2pt}%
  \fontfamily{#3}\fontseries{#4}\fontshape{#5}%
  \selectfont}%
\fi\endgroup%
\begin{picture}(8823,7101)(1093,-6109)
\put(6451,-2161){\makebox(0,0)[lb]{\smash{{\SetFigFont{25}{30.0}{\familydefault}{\mddefault}{\updefault}{\color[rgb]{0,0,0}$\vdots$}%
}}}}
\put(8101,-2161){\makebox(0,0)[lb]{\smash{{\SetFigFont{25}{30.0}{\familydefault}{\mddefault}{\updefault}{\color[rgb]{0,0,0}$\vdots$}%
}}}}
\put(4801,-2161){\makebox(0,0)[lb]{\smash{{\SetFigFont{25}{30.0}{\familydefault}{\mddefault}{\updefault}{\color[rgb]{0,0,0}$\vdots$}%
}}}}
\put(9901,-4411){\makebox(0,0)[lb]{\smash{{\SetFigFont{25}{30.0}{\familydefault}{\mddefault}{\updefault}{\color[rgb]{0,0,0}$Z_1$}%
}}}}
\put(9901,-3286){\makebox(0,0)[lb]{\smash{{\SetFigFont{25}{30.0}{\familydefault}{\mddefault}{\updefault}{\color[rgb]{0,0,0}$Z_2$}%
}}}}
\put(9901,-886){\makebox(0,0)[lb]{\smash{{\SetFigFont{25}{30.0}{\familydefault}{\mddefault}{\updefault}{\color[rgb]{0,0,0}$Z_{R-1}$}%
}}}}
\put(1801,-4786){\makebox(0,0)[lb]{\smash{{\SetFigFont{20}{24.0}{\familydefault}{\mddefault}{\updefault}{\color[rgb]{0,0,1}$\supp(\rho_2)$}%
}}}}
\put(6676,-5986){\makebox(0,0)[lb]{\smash{{\SetFigFont{20}{24.0}{\familydefault}{\mddefault}{\updefault}{\color[rgb]{0,0,1}$\supp(\rho_1)$}%
}}}}
\put(6226,-736){\makebox(0,0)[lb]{\smash{{\SetFigFont{20}{24.0}{\familydefault}{\mddefault}{\updefault}{\color[rgb]{0,0,1}$\supp(\rho_{R-1})$}%
}}}}
\put(2176,689){\makebox(0,0)[lb]{\smash{{\SetFigFont{20}{24.0}{\familydefault}{\mddefault}{\updefault}{\color[rgb]{0,0,1}$\supp(\rho_R)$}%
}}}}
\put(6001,-3811){\makebox(0,0)[lb]{\smash{{\SetFigFont{20}{24.0}{\familydefault}{\mddefault}{\updefault}{\color[rgb]{0,0,1}$\supp(\rho_3)$}%
}}}}
\put(9901,-5536){\makebox(0,0)[lb]{\smash{{\SetFigFont{25}{30.0}{\familydefault}{\mddefault}{\updefault}{\color[rgb]{0,0,0}$Z_0$}%
}}}}
\put(9901,464){\makebox(0,0)[lb]{\smash{{\SetFigFont{25}{30.0}{\familydefault}{\mddefault}{\updefault}{\color[rgb]{0,0,0}$Z_R$}%
}}}}
\end{picture}%

%% file: A0.pdf_t
\begin{picture}(0,0)%
\includegraphics{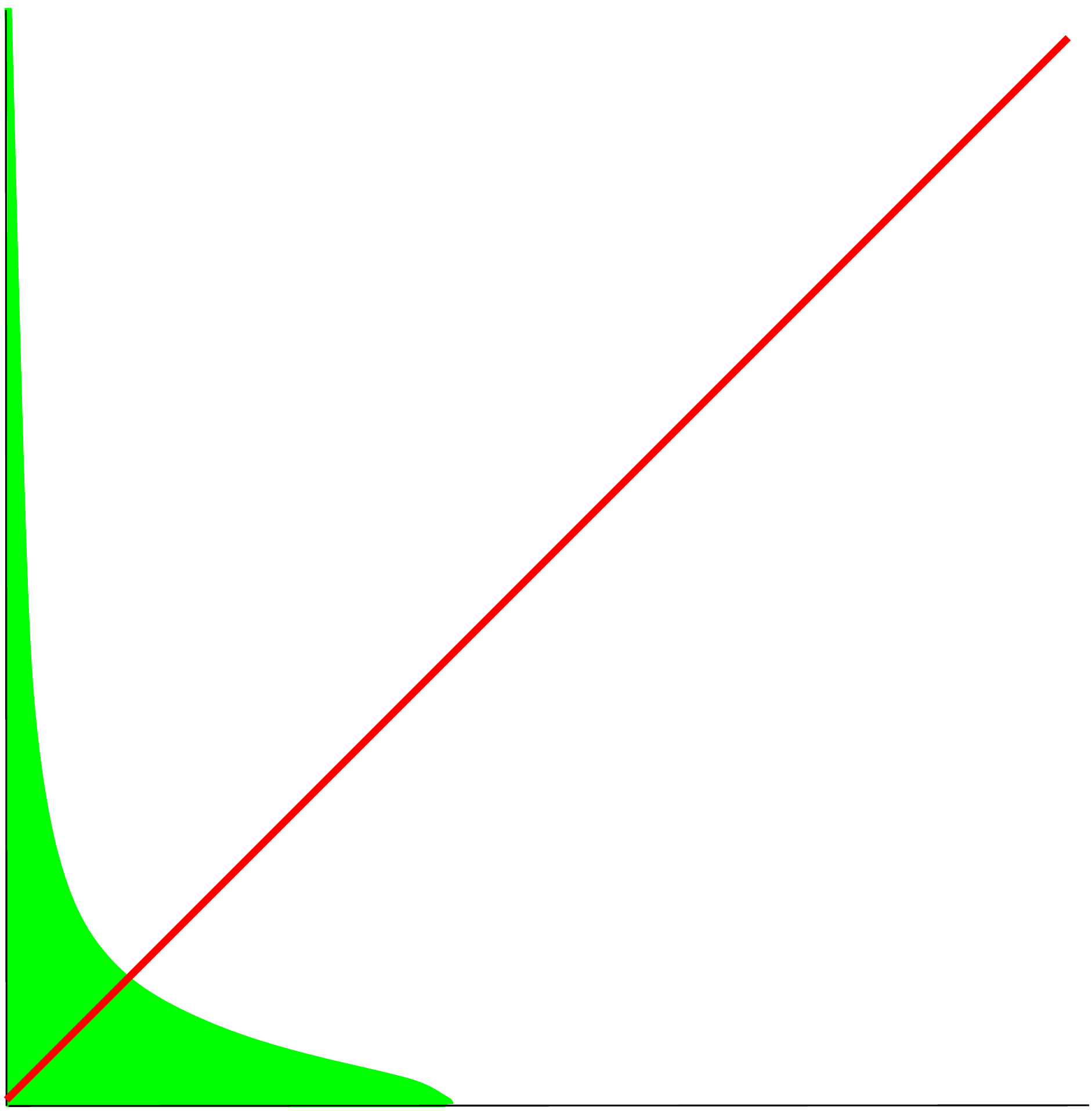}%
\end{picture}%
\setlength{\unitlength}{3947sp}%
\begingroup\makeatletter\ifx\SetFigFont\undefined%
\gdef\SetFigFont#1#2#3#4#5{%
  \reset@font\fontsize{#1}{#2pt}%
  \fontfamily{#3}\fontseries{#4}\fontshape{#5}%
  \selectfont}%
\fi\endgroup%
\begin{picture}(9033,9172)(2834,-11139)
\put(3845,-8201){\makebox(0,0)[lb]{\smash{{\SetFigFont{25}{30.0}{\familydefault}{\mddefault}{\updefault}{\color[rgb]{0,0,0}$a=0$}%
}}}}
\end{picture}%

%% file: A1.pdf_t
\begin{picture}(0,0)%
\includegraphics{A1.pdf}%
\end{picture}%
\setlength{\unitlength}{3947sp}%
\begingroup\makeatletter\ifx\SetFigFont\undefined%
\gdef\SetFigFont#1#2#3#4#5{%
  \reset@font\fontsize{#1}{#2pt}%
  \fontfamily{#3}\fontseries{#4}\fontshape{#5}%
  \selectfont}%
\fi\endgroup%
\begin{picture}(9030,9103)(2417,-13003)
\put(3475,-9941){\makebox(0,0)[lb]{\smash{{\SetFigFont{25}{30.0}{\familydefault}{\mddefault}{\updefault}{\color[rgb]{0,0,0}$a=1$}%
}}}}
\end{picture}%

%% file: A2.pdf_t
\begin{picture}(0,0)%
\includegraphics{A2.pdf}%
\end{picture}%
\setlength{\unitlength}{3947sp}%
\begingroup\makeatletter\ifx\SetFigFont\undefined%
\gdef\SetFigFont#1#2#3#4#5{%
  \reset@font\fontsize{#1}{#2pt}%
  \fontfamily{#3}\fontseries{#4}\fontshape{#5}%
  \selectfont}%
\fi\endgroup%
\begin{picture}(9094,9104)(2698,-15983)
\put(4245,-13021){\makebox(0,0)[lb]{\smash{{\SetFigFont{25}{30.0}{\familydefault}{\mddefault}{\updefault}{\color[rgb]{0,0,0}$a=2$}%
}}}}
\end{picture}%

%% file: A3.pdf_t
\begin{picture}(0,0)%
\includegraphics{A3.pdf}%
\end{picture}%
\setlength{\unitlength}{3947sp}%
\begingroup\makeatletter\ifx\SetFigFont\undefined%
\gdef\SetFigFont#1#2#3#4#5{%
  \reset@font\fontsize{#1}{#2pt}%
  \fontfamily{#3}\fontseries{#4}\fontshape{#5}%
  \selectfont}%
\fi\endgroup%
\begin{picture}(9043,9104)(4349,-16513)
\put(6225,-13261){\makebox(0,0)[lb]{\smash{{\SetFigFont{25}{30.0}{\familydefault}{\mddefault}{\updefault}{\color[rgb]{0,0,0}$a=3$}%
}}}}
\end{picture}%

%% file: Hurwitz3.pdf_t
\begin{picture}(0,0)%
\includegraphics{Hurwitz3.pdf}%
\end{picture}%
\setlength{\unitlength}{3947sp}%
\begingroup\makeatletter\ifx\SetFigFont\undefined%
\gdef\SetFigFont#1#2#3#4#5{%
  \reset@font\fontsize{#1}{#2pt}%
  \fontfamily{#3}\fontseries{#4}\fontshape{#5}%
  \selectfont}%
\fi\endgroup%
\begin{picture}(5673,3477)(2593,-6232)
\put(6901,-6136){\makebox(0,0)[lb]{\smash{{\SetFigFont{17}{20.4}{\familydefault}{\mddefault}{\updefault}{\color[rgb]{0,0,0}$3\sqrt{\frac 32}$}%
}}}}
\put(3001,-2986){\makebox(0,0)[lb]{\smash{{\SetFigFont{17}{20.4}{\familydefault}{\mddefault}{\updefault}{\color[rgb]{0,0,0}$-3\sqrt{\frac 32}$}%
}}}}
\put(5851,-5086){\makebox(0,0)[lb]{\smash{{\SetFigFont{17}{20.4}{\familydefault}{\mddefault}{\updefault}{\color[rgb]{0,0,1}$\supp(\rho_1)$}%
}}}}
\put(3976,-3961){\makebox(0,0)[lb]{\smash{{\SetFigFont{17}{20.4}{\familydefault}{\mddefault}{\updefault}{\color[rgb]{0,1,0}$\supp(\rho_2)$}%
}}}}
\put(8176,-3286){\makebox(0,0)[lb]{\smash{{\SetFigFont{25}{30.0}{\familydefault}{\mddefault}{\updefault}{\color[rgb]{0,0,0}$Z_2$}%
}}}}
\put(8251,-4411){\makebox(0,0)[lb]{\smash{{\SetFigFont{25}{30.0}{\familydefault}{\mddefault}{\updefault}{\color[rgb]{0,0,0}$Z_1$}%
}}}}
\put(8176,-5536){\makebox(0,0)[lb]{\smash{{\SetFigFont{25}{30.0}{\familydefault}{\mddefault}{\updefault}{\color[rgb]{0,0,0}$Z_0$}%
}}}}
\end{picture}%

%% file: EquilibriumMeasures_revision.bbl
\def\cprime{$'$} \def\cydot{\leavevmode\raise.4ex\hbox{.}}
\begin{thebibliography}{10}

\bibitem{Nikishin}
E.~M. Nikishin.
\newblock Simultaneous {P}ad\'e approximants.
\newblock {\em Mat. Sb. (N.S.)}, 113(155)(4(12)):499--519, 637, 1980.

\bibitem{GoncharRakhmanovPade}
A.~A. Gonchar and E.~A. Rakhmanov.
\newblock On the convergence of simultaneous {P}ad\'e approximants for systems
  of functions of {M}arkov type.
\newblock {\em Trudy Mat. Inst. Steklov.}, 157:31--48, 234, 1981.
\newblock Number theory, mathematical analysis and their applications.

\bibitem{GoncharRakhmanov}
A.~A. Gonchar and E.~A. Rakhmanov.
\newblock The equilibrium problem for vector potentials.
\newblock {\em Uspekhi Mat. Nauk}, 40(4(244)):155--156, 1985.

\bibitem{ns}
E.~M. Nikishin and V.~N. Sorokin.
\newblock {\em Rational approximations and orthogonality}, volume~92 of {\em
  Translations of Mathematical Monographs}.
\newblock American Mathematical Society, Providence, RI, 1991.
\newblock Translated from the Russian by Ralph P. Boas.

\bibitem{GoncharRakhmanovSorokin}
A.~A. Gonchar, E.~A. Rakhmanov, and V.~N. Sorokin.
\newblock On {H}ermite-{P}ad\'e approximants for systems of functions of
  {M}arkov type.
\newblock {\em Mat. Sb.}, 188(5):33--58, 1997.

\bibitem{Aptekarev}
A.~I. Aptekarev.
\newblock Strong asymptotics of polynomials of simultaneous orthogonality for
  {N}ikishin systems.
\newblock {\em Mat. Sb.}, 190(5):3--44, 1999.

\bibitem{Lapik}
M.~A. Lapik.
\newblock On the support of the extremal measure in a vector equilibrium
  problem.
\newblock {\em Mat. Sb.}, 197(8):101--118, 2006.

\bibitem{BertoGekhtSzmig1}
M.~Bertola, M.~Gekhtman, and J.~Szmigielski.
\newblock Peakons and {C}auchy {B}iorthogonal {P}olynomials.
\newblock {\em arXiv:0711.4082}, 2008.

\bibitem{dp}
A.~Degasperis and M.~Procesi.
\newblock Asymptotic integrability.
\newblock In A.~Degasperis and G.~Gaeta, editors, {\em Symmetry and
  perturbation theory (Rome, 1998)}, pages 23--37. World Scientific Publishing,
  River Edge, NJ, 1999.

\bibitem{BertoGekhtSzmig2}
M.~Bertola, M.~Gekhtman, and J.~Szmigielski.
\newblock The {C}auchy two--matrix model.
\newblock {\em arXiv:08040873}, 2008.

\bibitem{MehtaBook}
M.~L. Mehta.
\newblock {\em Random matrices}, volume 142 of {\em Pure and Applied
  Mathematics (Amsterdam)}.
\newblock Elsevier/Academic Press, Amsterdam, third edition, 2004.

\bibitem{VanAssche}
W.~Van~Assche, J.~S. Geronimo, and A.~B.~J. Kuijlaars.
\newblock Riemann-{H}ilbert problems for multiple orthogonal polynomials.
\newblock In {\em Special functions 2000: current perspective and future
  directions (Tempe, AZ)}, volume~30 of {\em NATO Sci. Ser. II Math. Phys.
  Chem.}, pages 23--59. Kluwer Acad. Publ., Dordrecht, 2001.

\bibitem{BertoGekhtSzmig3}
M.~Bertola, M.~Gekhtman, and J.~Szmigielski.
\newblock Strong asymptotics of {C}auchy biorthogonal polynomials.
\newblock {\em in preparation}, 2008.

\bibitem{DKMVZ}
P.~Deift, T.~Kriecherbauer, K.~T-R McLaughlin, S.~Venakides, and X.~Zhou.
\newblock Strong asymptotics of orthogonal polynomials with respect to
  exponential weights.
\newblock {\em Comm. Pure Appl. Math.}, 52(12):1491--1552, 1999.

\bibitem{Nuttall}
J.~Nuttall.
\newblock Asymptotics of diagonal {H}ermite-{P}ad\'e polynomials.
\newblock {\em J. Approx. Theory}, 42(4):299--386, 1984.

\bibitem{McLaughlinDeiftKriecherbauer}
P.~Deift, T.~Kriecherbauer, and K.~T.-R. McLaughlin.
\newblock New results on the equilibrium measure for logarithmic potentials in
  the presence of an external field.
\newblock {\em J. Approx. Theory}, 95(3):388--475, 1998.

\bibitem{Pastur}
L.~A. Pastur.
\newblock Spectral and probabilistic aspects of matrix models.
\newblock In {\em Algebraic and geometric methods in mathematical physics
  (Kaciveli, 1993)}, volume~19 of {\em Math. Phys. Stud.}, pages 207--242.
  Kluwer Acad. Publ., Dordrecht, 1996.

\bibitem{McLaughlin}
K.~T.-R. McLaughlin.
\newblock Asymptotic analysis of random matrices with external source and a
  family of algebraic curves.
\newblock {\em Nonlinearity}, 20(7):1547--1571, 2007.

\bibitem{Lysov}
V.~G. Lysov.
\newblock Strong asymptotics of {H}ermite-{P}ad\'e approximants for a system of
  {S}tieltjes functions with {L}aguerre weight.
\newblock {\em Mat. Sb.}, 196(12):99--122, 2005.

\bibitem{SaffTotik}
E.~B. Saff and V.~Totik.
\newblock {\em Logarithmic potentials with external fields}, volume 316 of {\em
  Grundlehren der Mathematischen Wissenschaften [Fundamental Principles of
  Mathematical Sciences]}.
\newblock Springer-Verlag, Berlin, 1997.
\newblock Appendix B by Thomas Bloom.

\bibitem{Prokhorov}
Yu.~V. Prohorov.
\newblock Convergence of random processes and limit theorems in probability
  theory.
\newblock {\em Teor. Veroyatnost. i Primenen.}, 1:177--238, 1956.

\end{thebibliography}
